# "Studies of Effects of Current on Exchange-Bias: A Brief Review",

J. Bass,[1] A. Sharma,[1] Z. Wei,[2] and M. Tsoi[2]
[1] Department of Physics and Astronomy, Michigan State University, East Lansing, MI 48824, USA
[2] Department of Physics, University of Texas at Austin, Austin, TX 78712, USA

MacDonald and co-workers recently predicted that high current densities could affect the magnetic order of antiferromagnetic (AFM) multilayers, in ways similar to those that occur in ferromagnetic (F) multilayers, and that changes in AFM magnetic order can produce an antiferromagnetic Giant Magnetoresistance (AGMR). Four groups have now studied current-driven effects on exchange bias at F/AFM interfaces. In this paper, we first briefly review the main predictions by MacDonald and co-workers, and then the results of experiments on exchange bias that these predictions stimulated.

## I. Introduction and Overview.

Spintronics in metallic multilayers composed of ferromagnetic (F) and non-magnetic (N) metals grew out of two complementary discoveries. The first, Giant Magnetoresistance (GMR)[1-6] , the discovery of which was awarded the 2007 Nobel Prize in Physics, involves a change in resistance of such multilayers when the relative orientation of the magnetic moments in adjacent F-layers is altered by an applied magnetic field H. The second, Spin-Transfer-Torque (STT) [7-12], involves a change in the relative orientations of the moments of the two F-layers in an F/N/F trilayer, driven by a large enough applied current density j. A question of fundamental importance is: Can similar effects occur in magnetic systems other than ferromagnets?

Recently, MacDonald and co-workers [13-16] predicted that they could occur in systems where the F-layers are replaced by antiferromagnetic (AFM) layers. First, they predicted that the resistance of an AFM trilayer of the form AFM/N/AFM could change when the relative orientation of the magnetic moments in the two AFM layers next to the N-layer changes--antiferromagnetic GMR = AGMR, and that the resistance changes could be comparable in size to those for GMR. Second, they predicted that injection of a large enough j, perpendicularly into an AFM/N/AFM trilayer (current-perpendicular-to-plane (CPP) geometry), could change the magnetic order of the trilayer. Their estimate of the necessary j $\sim 10^5$ A/cm$^2$ [13] was less than the typical j $\geq 10^7$ A/cm$^2$ needed to reverse the magnetic order in F/N/F multilayers [9-12,17]. Part of the reason for this smaller j is their conclusion that the STT with AFMs acts on a large portion of the AFM-metal, whereas it acts on an F-metal only near the N/F interface. Last, they predicted that a large enough CPP j injected into an F/AFM interface could affect the exchange-bias at the interface [14]. Such a phenomenon could allow current control of exchange-bias in magnetic devices. Replacing F-metals in spintronic devices with AFM-metals would also eliminate unwanted effects of shape anisotropy on the magnetic stability of small elements, thus potentially offering better control of the magnetic state in nanoscale systems. Xu et al. [18] recently calculated the AGMR for a simple AFM/N/AFM/N = FeMn/Cu/FeMn/Cu multilayer, and found results similar to those predicted by MacDonald and co-workers, and Gomonay and Loktev [19] provided additional theoretical evidence that polarized current can destabilize the equilibrium state of an AFM.

Importantly for our present purposes, all of the published calculations are for perfect layers and ballistic transport. The STT and resulting AGMR are consequences of quantum interference. The authors note that disorder that produces diffusive scattering, and weakens quantum interference, will weaken any such STT and AGMR.

Given that the samples studied so far are imperfect, with transport in both F and AFM layers, and probably also at AFM/N and AFM/F interfaces, that is significantly diffusive, these calculations are likely to give only qualitative guidance to interpreting experiments on such samples. We, thus, view the experimental results as providing as much guidance to theory as vice-versa.

Stimulated by the theoretical studies listed above, four experimental searches for effects of spin-torque on AFMs have been published, all working with exchange-biased spin-valves (EBSVs) of the form AFM/F2/N/F1 [20-23]. Here the AFM lies outside the 'active' GMR region of the two F-layers, and serves mainly to 'pin' the magnetization of the adjacent F2-layer to a higher reversing (switching) field than that of the 'free' F1 layer, leaving the F1 layer free to rotate at a lower field. The pinning is produced either by heating the sample to above the blocking temperature of the AFM, applying a magnetic field, and then cooling to room temperature with the field on, or else by applying a magnetic field during sample growth. We define the exchange bias field as $H_E = -(H^L + H^R)/2$ and the coercive field as $H_C = -(H^L – H^R)/2$, where $H^L$ is the leftmost field at which the pinned layer flips, and $H^R$ is the rightmost field. With the standard definitions of field directions, $H^L$ is more negative than $H^R$, so both $H_E$ and $H_C$ are positive. Fig. 1 shows examples of the major hysteresis loops for three different classes of EBSVs: (A) $H_E > H_C$, (B) $H_E \approx H_C$, and (C) $H_E < H_C$. In all three cases, we take the 'free' and 'pinned' layers to be magnetically uncoupled, so that the 'free' layer reverses symmetrically about H = 0. We start with large + H along the pinning direction, causing the moments of both the 'free' and 'pinned' layers to point along +H (parallel—P) orientation. From standard GMR theory with identical F-metals, $R_P$ is the minimum resistance. Reducing the magnitude of H, the 'free' layer reverses at a small magnitude negative H, giving maximum resistance, $R_{AP}$. A larger magnitude negative $H^L$ is needed to break the EB pinning of the 'pinned' layer, at which its moment rotates to along –H, returning the sample to $R_P$. After $R_P$ is achieved, the direction of change of H is reversed and the field is swept back toward +H. In case A, $H_E > H_C$, the 'pinned' layer switches back to its preferred pinned direction (at $H^R$) before reaching H = 0, giving $R_{AP}$. The sample returns to $R_P$ after passing through H = 0, when the free layer reverses. In case B, $H_E \approx H_C$, the reversal of the pinned layer doesn't occur until $H^R \approx 0$, and R may never reach $R_{AP}$ before falling back to $R_P$ when the 'free' layer flips. In case C, $|H_E| < |H_C|$, the 'pinned' layer doesn't unpin until $H^R$ is beyond where the free layer flips.

Case A is seen in study [20]. Case C is seen in studies [21] and [23]. Something closest to case B looks to be seen in study [22], where the switching of all layers before reaching H = 0 indicates that the two F-layers are magnetically coupled. F2 is then not completely 'free', but is driven parallel to F1 soon after F1 starts to flip back to the +H direction.

In addition to their different classes of EBSVs just described, the four published studies of effects of spin-torque on EBSVs also have other differences. Wei et al. [20] used a point contact to inject a high, dc, approximately CPP, current density, j ~ $10^8$ A/cm$^2$ , into an EBSV film. Urazhdin and Anthony [21], sent a dc CPP current density j ~ 5 x $10^7$ A/cm$^2$ into nanopillar EBSVs. Tang et al.[22] sent a dc current-in-plane (CIP) current density ~ $10^6$ A/cm$^2$ (The $10^6$ A/m$^2$ specified in [22] was a misprint [24]) into an EBSV film with a metallic AFM. Nam et al.

[23] sent an ac CIP current density j ~ $10^5$ A/cm$^2$ into an EBSV film with an insulating AFM. In the following we describe the results reported in each study.

**II. CPP study of effect of dc *I* upon exchange bias using adjustable point contacts.**

Wei et al. [20] measured point contact resistances at room temperature (~ 295K) with negative current flowing from the contact tip into the sample. The sample geometry is shown in Fig. 2. A point contact is used to inject a dc current into a sputtered F1/N/F2/AFM/N = Cu(50 or 100)/CoFe(3 or 10)/Cu(10)/CoFe(3 or 10)/FeMn(3 or 8)/Au(5) multilayer (or inverted versions thereof—i.e. Cu/ FeMn/CoFe/Cu/CoFe/Au). Here, and hereafter, layer thicknesses are given in nm, and CoFe = $Co_{91}Fe_9$. The sample is heated to ~ 450K (above the blocking temperature of FeMn), and then cooled in a magnetic field of 180 Oe to exchange bias the 'pinned' layer F2 to a higher magnetic field than needed to reverse the 'free' layer F2. To protect the top layer from atmospheric contamination, it is covered by a 5 nm thick layer of Au. The magnetic field H is applied in the plane of the layers and along the direction of exchange-bias. Magnetic coupling between the two F-layers should be negligible, because the N layer is thick enough (10 nm) to eliminate exchange coupling, and the two F-layers are wide enough (~ mm) to minimize dipolar coupling. The bottom N layer is Cu, made thick enough (50 or 100 nm) to approximate an equipotential, thereby generating an approximately CPP current flow through the F1/N/F2/AFM EBSV. Fig. 3 [20] shows 'type A' switching curves (Fig. 1) for a sample with thicknesses F1 = 10 nm, F2 = 3 nm, and FeMn = 8 nm for a range of both positive and negative currents *I*. For a point contact with resistance R = 0.92 Ω, the dark curves show sweeps from positive to negative field, and the lighter curves show sweeps back down from negative to positive field. For this contact, *I* = 30 mA corresponds to j ~ 2 x $10^8$ A/cm$^2$.

Focusing upon the dark curves, we see that the switching field of the free layer F1 is essentially independent of the magnitude of *I* and shows little broadening. In contrast, the switching field of the pinned layer F2 broadens significantly as the magnitude of *I* increases, and the midpoint of the switching also shifts with *I*, increasing for – *I* and decreasing for + *I*. Similar behaviors are seen also in the lighter curves. Opposite shifts for + *I* and – *I* indicate that these shifts cannot be due to Joule heating, which should give shifts in the same direction for both directions of *I*. But Joule heating might contribute to the broadening of the switching transitions. The shifts of the dark curves are specified more clearly in Fig. 4a [20], which shows grey-scale plots of the heights of the curves in Fig. 3, with white representing the anti-parallel (AP) state of maximum resistance, and black the parallel (P) state of minimum resistance. Data for three representative contacts (out of 29), on three different samples, show that the behavior of interest is not limited to a single sample or contact, and that similar results are generally obtained for straight line fits to 30% (white dashed lines), 50% (solid white lines), and 70% (solid black lines) of the maximum change in R. The first sample in Fig. 4a is the one from Fig. 3. The second is a contact with R = 1.6 Ω to a similar sample, but with the AFM layer on top (i.e. closest to the point contact), so that the 'directions' of currents are reversed. The third has the AFM layer back on the bottom, but equal thickness F1 = F2 = 3 nm layers. All three samples show the same features—i.e. electrons passing through F2 into the AFM-layer enhance pinning, and electrons passing through the AFM-layer into F1 reduce it.

The data in Figs. 3 and 4 show that similar linear variations occur with both up and down sweeps of H. Thus $H_E$ increases with increasing magnitude of - *I* and decreases with increasing + *I*. Wei et al. proposed a qualitative explanation for these asymmetric changes in switching field and $H_E$ with *I*. Their model is shown schematically in Fig. 5 [20]. The AFM is assumed to be ‘uncompensated’—i.e. composed of individual layers that are F-like in the plane defined by the interface, but with moments in adjacent layers alternating ‘up’ or ‘down’ to give net magnetization M = 0. The AFM layers are divided into two parts: (a) ‘bulk’ layers, in blue—with an assumed AFM ‘domain’ indicated by a dotted boundary, and (b) an interfacial layer indicated in grey. The latter is assumed to consist of two parts. First, some ‘fixed’ moments that lie parallel to the F/AFM interface and do not change orientation with either H or I. These moments give the pinning at $I = 0$. Second, some ‘free’ moments that are strongly exchange coupled to the ‘bulk domain’, and thus rotate with it. Following the ideas of MacDonald and coworkers, the current is assumed to exert a STT on the AFM ‘domain’, rotating it toward the direction of the F-moments (- *I*) or away from this direction (+ *I*). The resulting rotation of the ‘free’ interfacial moments increases the pinning for – *I* and decreases it for + *I*, as observed in Figs. 3 and 4.

**III. CPP study of effect of dc *I* upon exchange bias using nanopillars.**

Urazhdin and Anthony [21] studied the effect of a dc CPP current at 4.2K on sputtered and ion-milled 120 nm x 60 nm nanopillar EBSVs of the form Py(30)/Cu(10)/Py(5)/FeMn(t)/Cu(1)/Au(10), ion-milled through part of the 10 nm thick middle Cu layer to leave the rest of that layer and the whole bottom (30 nm) Py layer extended in area. The FeMn thickness varied from t = 1.5 to 4 nm, and they focused on a sample with t = 1.5 nm. Positive current flowed from the extended to the patterned Py layer, and H was applied along the nanopillar easy axis. The current densities at which they found effects were j ~ 5 x $10^7$ A/cm$^2$.

They first looked for effects of *I* upon the AFM by applying a series of positive or negative pulse currents, $I_o$, at fields $H_o = \pm 3$ kOe (large enough to suppress current-induced reversal of the Py(5) layer), and then measuring $H_E$ at *I* small enough to not affect the magnetic state of the nanopillar. They found the effect of $I_o$ upon $H_E$ to be asymmetric in $H_o$: for negative $H_o$, they found initial linear increases of $H_E$ with the magnitude of $I_o$, then slight drops, and finally approximate saturation; for positive $H_o$, they found no systematic variation of $H_E$ with $I_o$ to within uncertainties. They took the eventual saturation of $H_E$ with magnitude of $I_o$, and the difference in behaviors between $+H_o$ and $–H_o$ as evidence that the behaviors for $–H_o$ were not caused by sample heating. They found the sharp increases in $H_E$ with $I_o$ to correlate with the presence of bipolar steps in differential resistance. They attributed the observed asymmetric behaviors of $H_E$ to spin-transfer torque (STT) acting on the Fe moments at the Py/FeMn interfaces, to asymmetrically enhance or suppress $H_E$, and to cause to precess the stable Fe moments that stay parallel to the Py magnetization, thereby inducing an indirect dynamical Py response. They tentatively attributed the symmetric enhancement of $H_E$ to a combination of FeMn moment rotations at the Py/FeMn interface stimulated by electron-magnon scattering at the interface, and/or Fe moment rotations engendered by a torque due to the Oersted field of the injected current. They also found deviations from the statistics expected for thermally activated switching of EBSVs, which they attributed to current-induced effects on magnetic layer fluctuations.

**IV. CIP Study of effect of dc *I* upon exchange-bias using multilayers.**

Tang et al. [22] injected a dc CIP current at 295K into 5 x 5 mm$^2$ sputtered, EBSV multilayers of the form Ta(10)/NiFe(10)/Cu(4)/NiFe(10)/FeMn(15)/Ta(5). The top NiFe layer was EB pinned to the FeMn layer by growing the sample in a field of ~ 300 Oe oriented parallel to the layers. CIP currents up to 220 mA were applied along or opposite to the pinning direction, and the magnetic field was swept 'up' or 'down' Examination of their hysteresis curves shows that the two F-layers are not magnetically uncoupled, since both layers 'reverse' before the magnetic field returns to H = 0.

The authors of [22] defined two 'states' of their sample. In state I, the 'pinned' direction was along + H, giving unpinning at negative H as in Fig. 1. As already noted, the occurrence of all switchings at negative H indicates some magnetic coupling between F1 and F2. In state II, the sample was rotated a nominal $180^o$ [24]. As expected, the switchings now all occurred at positive fields. Their main reported results were as follows. (1) The switching curves for states I and II at different values of applied $I_o$ were mostly only slightly different, except for $I_o > 150$ mA, when the MR disappeared more rapidly in state II with increasing $I_o$. Whether this difference is significant, or due to slight misalignments of the pinning direction, applied H, and applied $I_{o,}$ is not clear. (2) Increasing $I_o$ substantially decreased the magnitude of the pinned layer reversal field, $H^L$, from about 45 Oe at $I_o = 1$ mA to below 20 Oe by $I_o = 220$ mA. (3) The behavior of $H^R$ was less clear. Assuming a type (C) EBSV as in Fig. 1, the part of the hysteresis curves showing the start of the return of the pinned layer to its pinned direction for $I_o \leq 100$ mA could be interpreted as indicating no change in $H^R$ or the reversal field of the free layer. However, the decrease in height of the 'return peak' with increasing $I_o$, and its disappearance above $I_o = 100$ mA, might also mean that $H^R$ is becoming more negative, and rotating the coupled 'free' layer along with it at increasingly negative fields. Both might then reverse so closely together that their reversal gives no MR. (4) When the sample was in state II (i.e., rotated by $180^o$), and an 80 Oe field was applied opposite to the pinned direction [24], pulsing a 250 mA current for 1 s caused the hysteresis curve to reverse. No reversal was found when the sample started in state I and the 80 Oe field was applied along the pinned direction during the pulse [24]. The authors attributed their observed behaviors to a combination of effects due to the magnetic field produced by the large current *I* and current-induced torques on the AFM, in case (4) being effective only when the sample is initially in state II. A possible alternative explanation for case (4) is that the pulse heated the sample enough so that the 80 Oe field applied opposite to the pinning direction in state II reversed the pinning, whereas the same field applied along the pinning direction in state I left the pinning unchanged..

**V. CIP Study of effect of ac *I* upon exchange-bias using multilayers.**

Nam et al. [23] injected ac CIP current at 300K into a sputtered, exchange-biased 1 x 5 $mm^2$ multilayer of the form FeCo(2.4)/NiCoO(40)/FeCo(3)/Cu(3.8)/FeCo(4.5), with FeCo = $Fe_{15}Co_{85}$ and NiCoO = $Ni_{0.85}Co_{0.15}O$. The thinner FeCo layer was exchange-biased to the NiCoO layer by growth in a 500 Oe field aligned in the layer plane. R and MR were measured at frequency $f$ = 7.5 Hz with ac currents *I* ranging up to 90 mA set to flow only during the duration time $t_d$, fixed for standard measurements at $t_d = 0.3$ sec. H and *I* were aligned along the exchange-bias axis. NiCoO was chosen as a very high resistance AFM insulator, to ensure that no current flowed through it, thereby eliminating the possibility of spin-transfer-torque within the AFM.

As in prior studies, they found no significant effect of *I* on the switching field of the 'free' FeCo layer. And as in ref. [22], they found that increasing *I* decreased the magnitude of $H^L$, the large, negative H switching field of the pinned FeCo layer. However, they found that the magnitude of $H^R$, the smaller positive switching field of the pinned layer (case C in Fig. 1), also decreased with increasing *I*, so that $H_E$ stayed roughly constant up to $I = 60$ mA, corresponding to j ~ 5 x $10^5$ A/$cm^2$. They found that they could approximately reproduce their observed changes in R with *I* by increasing the sample temperature, with a change in temperature by $60^oC$ corresponding roughly to a change in current of 90 mA. But more detailed analysis showed that changes in resistance (and exchange bias) produced by current and temperature were not identical. They concluded that an additional mechanism beyond heating was also needed, and suggested the possibility of spin-transfer between the two F-layers.

## VI. Summary and Conclusions.

Predictions that strong current densities could produce spin-transfer-torque (STT)-like effects on antiferromagnets (AFMs) [13-16,18,19], stimulated four attempts to observe such effects in exchange-biased spin-valves (EBSVs). The calculations are all for perfect, single crystal samples, with the predictions requiring ballistic transport and quantum interference. In contrast, real sputtered samples are 'dirty' and transport is at least substantially diffusive. The calculations can, thus, give only qualitative guidance, and experiments are crucial to see if STT effects on AFMs appear in real sputtered samples.

Two somewhat different studies in the current-perpendicular-to-plane (CPP) geometry [20,21] produced evidence that current densities $j \geq 10^7$ A/cm$^2$ could affect the F/AFM interface, including the exchange-bias field ($H_E$). Ref. [20] found that one polarity of $I$ increased $H_E$, while the other polarity decreased it. Ref. [21] found opposite currents to produce roughly similar effects for a given direction of a large magnetic field, but different effects for opposite directions of the field. Ref. [20] proposed a model based upon static STT effects of the current on the bulk of the AFM. Ref. [21] proposed a combination of static and dynamic STT effects at the F/AFM interface.

Two studies in the current-in-plane (CIP) geometry leave the issue of effects of current on $H_E$ still unresolved. As corrected [24], Ref. [22] reported a decrease in $H^L$ due to dc currents $j \sim 10^6$ A/cm$^2$, and inferred a decrease in $H_E$, which, however, is less sure. In contrast, using ac currents as large as $2 \times 10^5$ A/cm$^2$, Ref. [23] reported similar decreases in $H^L$ to those in [22], but also decreases in $H^R$ similar to those in $H^L$, leading to no significant change in $H_E$ up to $I = 60$ mA. They showed that they could simulate most of the changes in their exchange-bias hysteresis curves simply by heating their sample.

We conclude that our understanding of effects of large $I$ upon exchange-bias is still very incomplete.

Acknowledgments. We thank X.-L. Tang for providing additional information about her experiments and S. Urazhdin for useful suggestions. This work was supported in part by the US NSF under grants DMR 05-01013 and DMR-06-45377.

## References

1. M.N. Baibich et al., Phys. Rev. Lett. **61**, 2472 (1988).
2. G. Binash et al., Phys. Rev. **B39**, 4828 (1989).
3. P.M. Levy, Solid State Physics Series, **47**, 367 (1994). H. Ehrenreich, D. Turnbull (Eds.), Academic Press, NY.
4. M.A.M. Gijs and G.E.W. Bauer, Adv. In Phys. **46**, 285 (1997).
5. J. Bass and W.P. Pratt Jr., J. Magn. Magn. Mat. **200**, 274 (1999).
6. A. Fert and L. Piraux, J. Magn. Magn. Mat. **200**, 338 (1999).
7. J. Slonczewski, J. Magn. Magn. Mat. **159**, L1 (1996).
8. L. Berger, J. Appl. Phys. **81**, 4880 (1997).
9. M. Tsoi et al., Phys. Rev. Lett. **80**, 4281 (1998).
10. E.B. Myers et al., Science **285**, 867 (1999).
11. J.Z. Sun, J. Magn. Magn. Mat. **202**, 157 (1999).
12. J.-E. Wegrow et al., Europhys. Lett. **45**, 626 (1999).
13. A.S. Nunez et al., ArXiv: cond-mat/0510797.
14. A.S. Nunez et al., Phys. Rev. **B73**, 214426 (2006).
15. P.M. Haney and A.H. MacDonald,ArXiv:0708.3231.
16. P.M. Haney et al., arXiv:0709.3862.
17. S. Urazhdin et al., Phys. Rev. Lett. **91**, 146803 (2003).
18. Y. Xu et al., arXiv: 0708.2143.
19. H.V. Gomonay and V.M. Loktev, arXiv:0709.4172.
20. Z. Wei et al., Phys. Rev. Lett. **98**, 116603 (2007).
21. S. Urazhdin and N. Anthony, Phys. Rev. Lett. **99**, 046602 (2007).
22. X.-L.Tang et al., Appl. Phys. Lett. **91**, 122504 (2007).
23. D.N.H. Nam et al., arXiv:0801.1515 (2008).
24. X.-L. Tang (Private Communication).

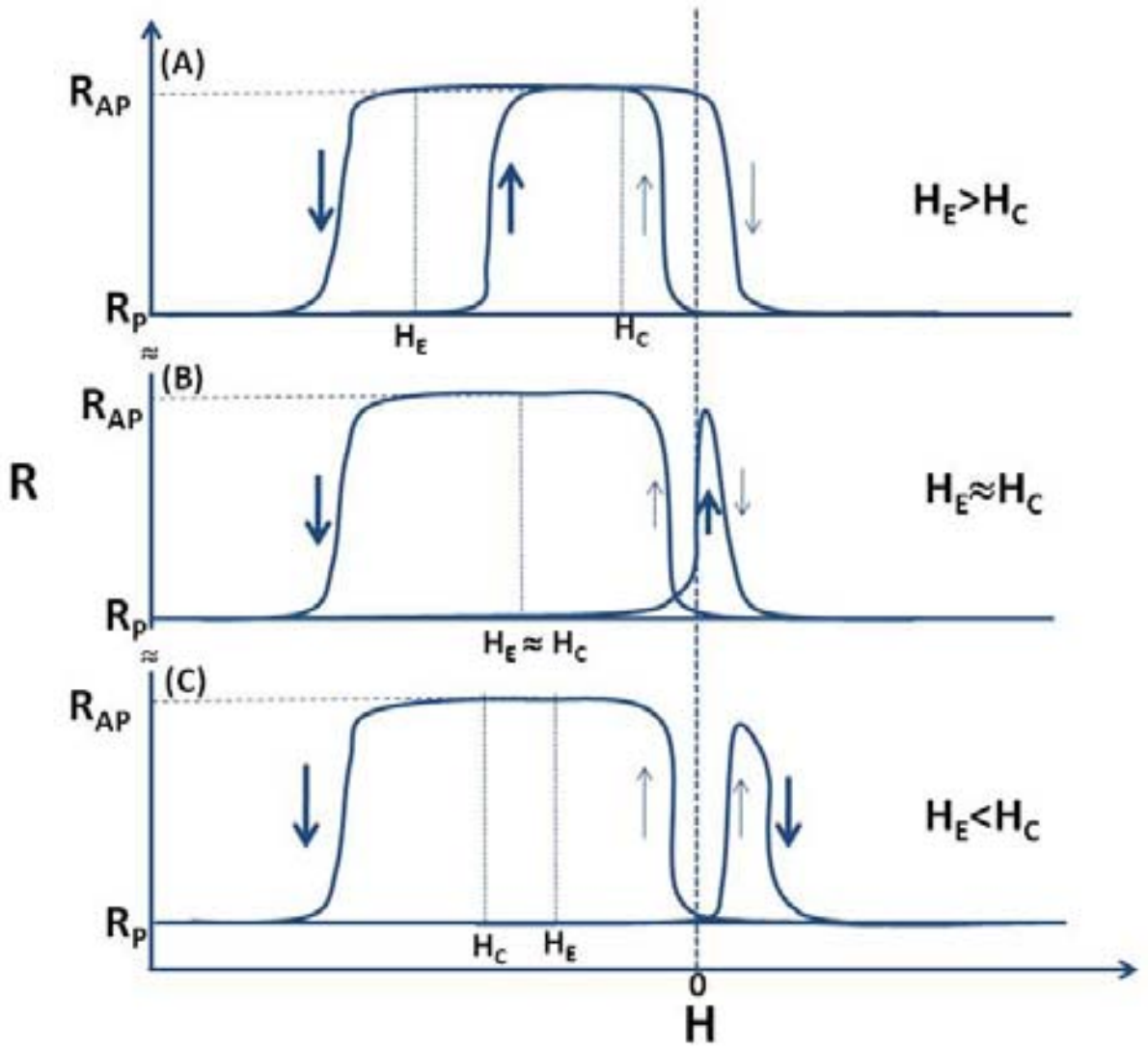

Fig. 1. Schematic hysteresis curves for exchange-biased spin-valves with: (A) $H_E > H_C$; (B) $H_E \sim H_C$, and (C) $H_E < H_C$. Thick arrows indicate when the pinned layer reverses; thin arrows indicate when the free layer reverses. The arrows point in the direction of resistance change. The free layer reverses symmetrically about H = 0.

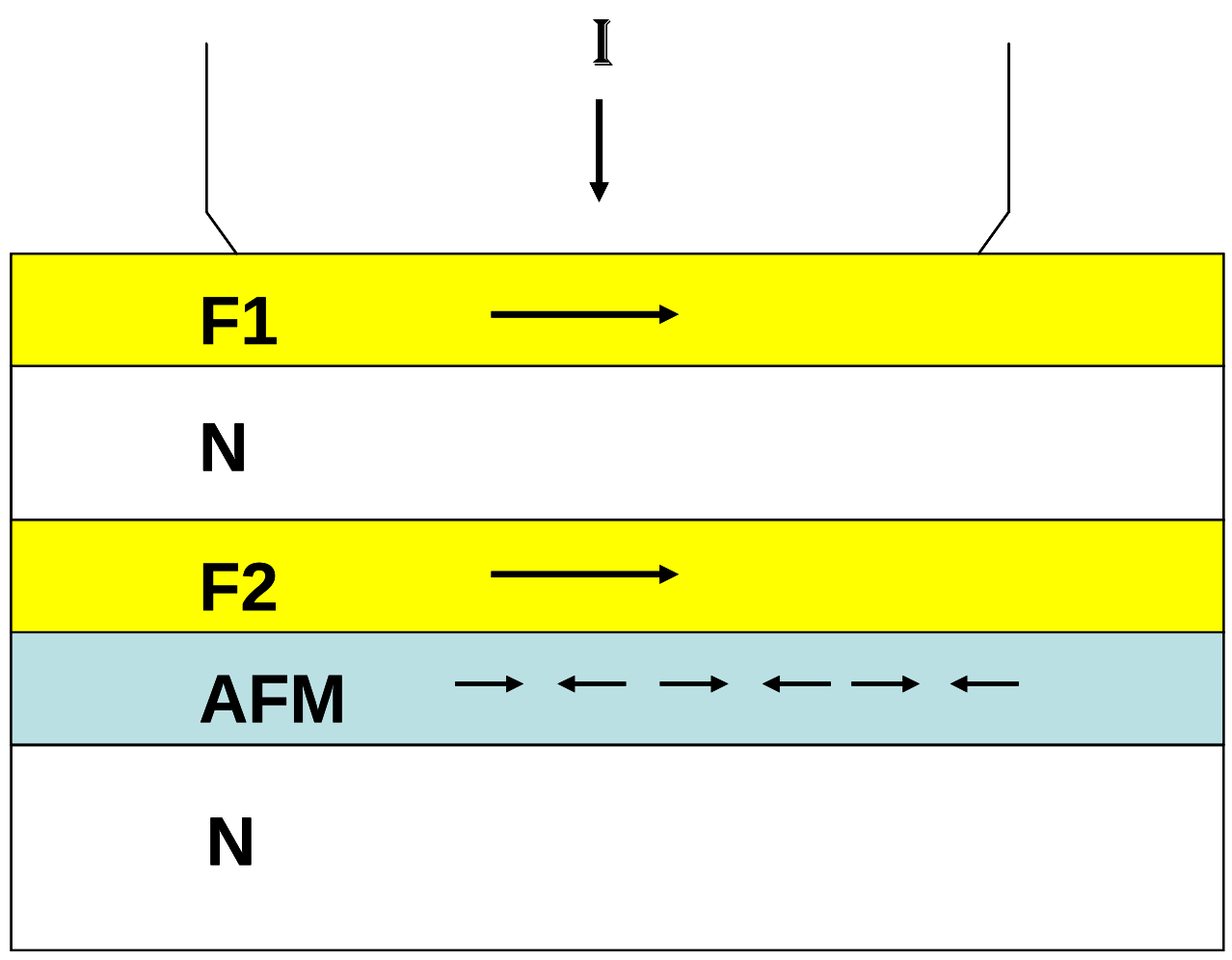

Fig. 2. Schematic of point contact to sample geometry. Omitted is a 5 nm thick protective Au capping layer between the point contact and the multilayer. The bottom N layer is also much thicker than shown, to help produce a nearly CPP current.

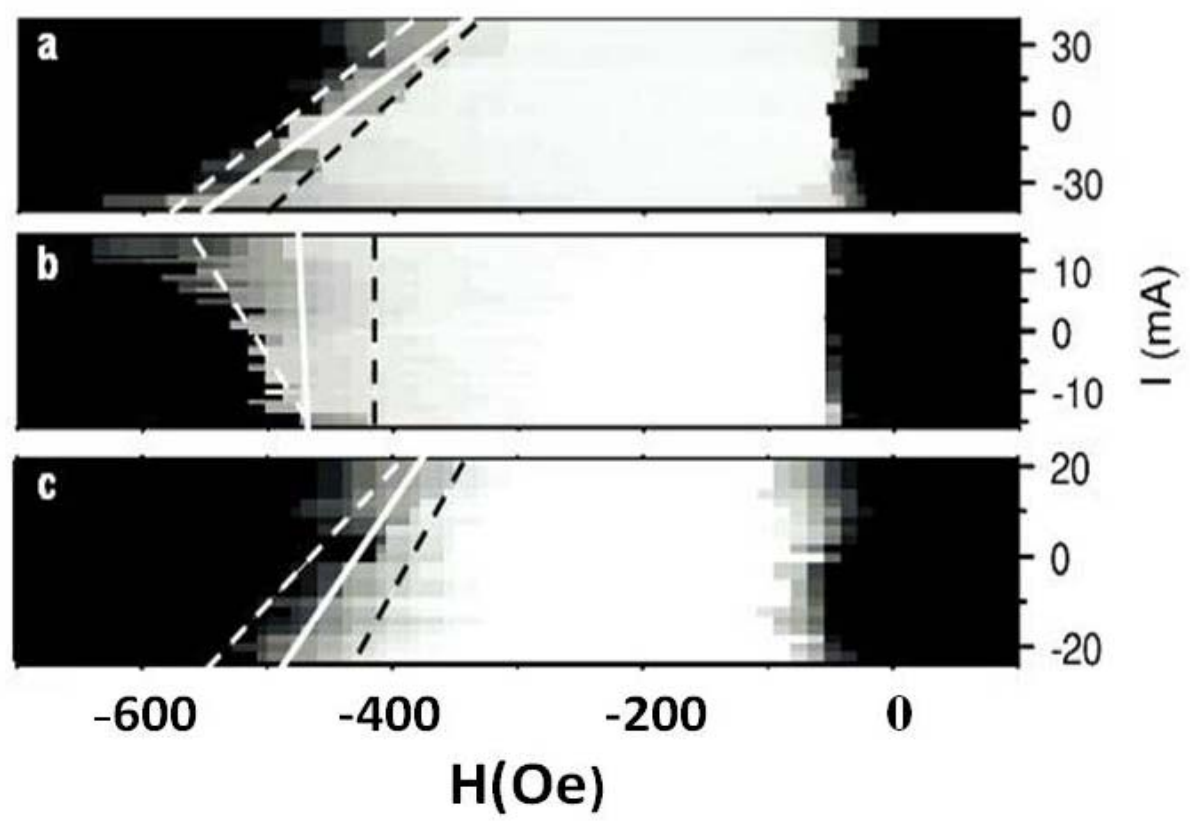

Fig. 4. Grey scale plots of resistance vs H for different values of I. White is maximum R (AP state) and Black is minimum R (P state). Lines are linear fits to the data at 30% (dashed white), 50% (solid white) and 70% (dashed black) of maximum R. Sample a is the sample of Fig. 2, with the AFM layer on the bottom (furthest from the point contact). Sample b is similar to a, except inverted, so that the AFM layer is on top (closest to the point contact). Sample c differs from a only in that it has two equally thick (8 nm) F-layers. For sample a, the field was swept down from high positive values (as for the black curves in Fig. 3). For samples b and c, the field was swept up from high negative values (as for the grey curves in Fig. 3). Figure after ref. [20].

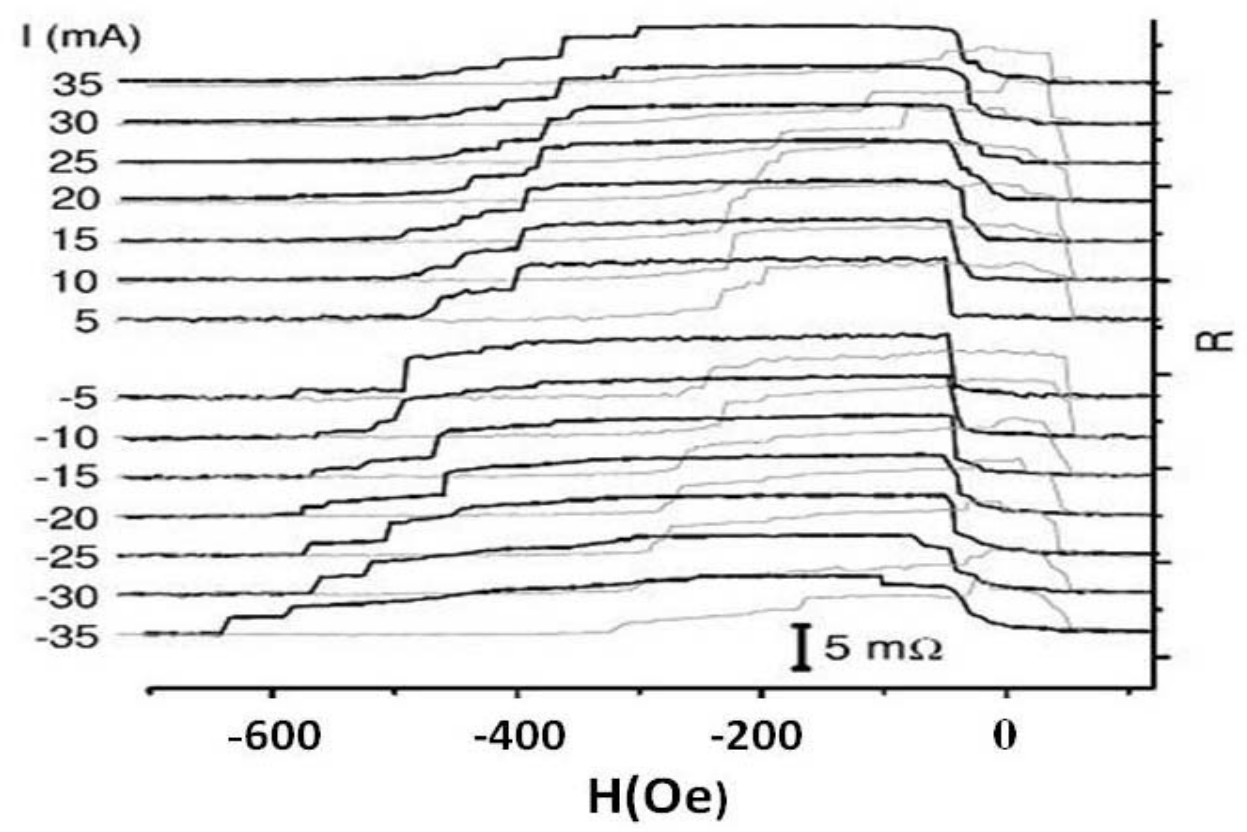

Fig. 3. R (vertical scale) vs magnetic field H for a series of currents *I*. The solid black curves are hysteresis curves starting from large positive H and finishing at large negative H. The grey curves start at large negative H and finish at positive H. In the dark curves, the 'free'layer, F1, switches at H ~ -50 Oe and the pinned layer, F2, switches at fields ranging from below -400 Oe (large, positive *I*) to about - 600 Oe (large magnitudes of negative *I*). Figure after ref. [20].

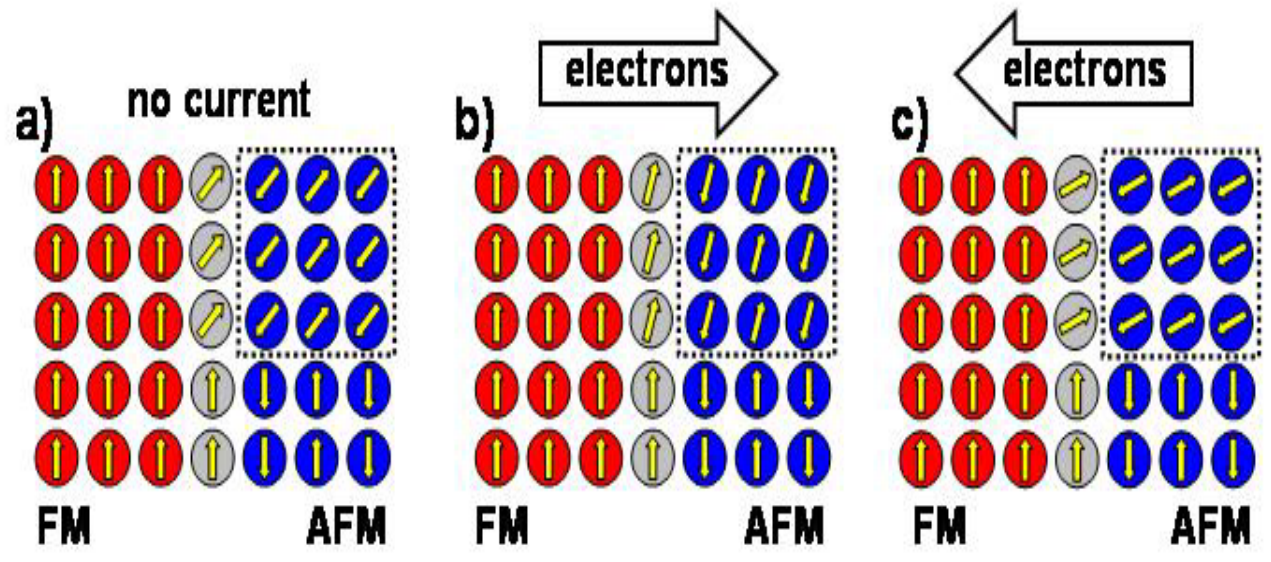

Fig. 5. Schematic model. Red is the F-metal. Grey are the AFM layer next to the F-metal. Vertical grey arrows are fixed moments that give the exchange-bias at I = 0. Tilted grey arrows are 'free' moments exchange coupled to the blue AFM domain indicated by the dotted box. As current rotates this domain, the free grey moments rotate with it. Electrons going from F to AFM rotate the AFM domain toward the vertical, increasing the exchange-bias. Electrons going from AFM to F rotate the AFM domain away from vertical, decreasing the exchange-bias. Figure from ref. [20].